\documentclass[referee]{aa}

\usepackage{graphicx}
\usepackage{natbib}
\usepackage{latexsym}
\usepackage{amsmath}
\usepackage{amssymb}
\usepackage{amstext}
\usepackage{color}
\usepackage{psfrag}

\def\k{km s$^{-1}$}
\def\ks{km s$^{-1}$~}

\def\m{$^\prime$}
\def\s{$^{\prime\prime}$}

\def\cm3{cm$^{-3}$}

\def\2{$^{12}$CO}
\def\3{$^{13}$CO}
\def\msol{M$_\odot$}

\begin{document}

\title{Identifying the counterpart of HESS J1858+020}

\author {S. Paron \inst{1}
 \and E. Giacani \inst{1}
}
                                                                                                               
\institute{Instituto de Astronom\'{\i}a y F\'{\i}sica del Espacio (CONICET-UBA),
             CC 67, Suc. 28, 1428 Buenos Aires, Argentina\\
             \email{sparon@iafe.uba.ar}
}

\offprints{S. Paron}

   \date{Received <date>; Accepted <date>}

\abstract{}{HESS J1858+020 is a weak $\gamma-$ray source that does not have any clear cataloged counterpart
at any wavelengths. Recently, the source G35.6-0.4 was re-identified as a SNR. The HESS source lies towards the 
southern border of this remnant. The purpose of this work is to investigate the interstellar medium around 
the mentioned sources in order to look for possible counterparts of the very-high energy emission.}{Using the \3 
J=1--0 line from the Galactic Ring Survey and mid-IR data from GLIMPSE we analyze the environs of 
HESS J1858+020 and SNR G35.6-0.4.}{The \3 data show the presence of a molecular cloud towards the southern 
border of SNR G35.6-0.4 and at the same distance as the remnant. This cloud is composed by two molecular clumps, 
one, over the SNR shell and the other located at the center of HESS J1858+020. We estimate a molecular mass 
and a density of $\sim 5 \times 10^{3}$ \msol~and $\sim 500$ cm$^{-3}$, respectively for
each clump. Considering the gamma-ray flux observed towards HESS J1858+020,
we estimate that a molecular cloud with a density of at least 150 cm$^{-3}$ could explain the
very-high energy emission hadronically. Thus, we suggest that the $\gamma-$ray emission detected 
in HESS J1858+020 is due to hadronic mechanism. Additionally, analyzing mid-IR emission, we find that the region is 
active in star formation, which could be considered as an alternative or complementary possibility to explain 
the very-high energy emission.}{}

\titlerunning{Indentifying the counterpart of HESS J1858+020}
\authorrunning{S. Paron et al.}

\keywords{ISM: clouds - ISM: supernova remnants - gamma rays: observations - ISM: individual objects: HESS J1858+020}

\maketitle

\section{Introduction}

Among the rich population of TeV $\gamma-$ray sources, a good fraction
are associated with Galactic phenomena which include supernova remnants (SNRs), 
pulsar wind nebulae (PWNe) and binary systems \citep{hinton08}. Extragalactic
TeV sources are associated with active AGNs (mostly blazars). There are, 
however, a number of $\gamma-$ray sources for which there are not yet a 
clear identification and an origin established. In this paper we focus on 
the very-high energy source HESS J1858+020. 

HESS J1858+020 is a weak $\gamma-$ray source detected with H.E.S.S. (the High
Energy Stereoscopic System), an array of air Cherenkov telescopes. 
Though nearly point-like 
source, its morphology shows a slight extension of $\sim 5^{\prime}$ 
along its major axis. The source has been detected at a significance level 
of 7$\sigma$ with a differential spectral index of 2.2 $\pm$0.1.
At the present, HESS J1858+020 does not have any clear cataloged counterpart 
at any wavelengths.  
The pulsar PSR J1858+0143, which is energetic enough to power HESS J1858+020, is located 
far away from the center of the gamma source, thus an association between them is 
unlikely \citep{aha08}.

Recently \citet{green09}, using radio continuum data from the Very Large Array (VLA) 
Galactic Plane Survey (VGPS; \citealt{stil06}), re-identified the radio continuum source
G35.6-0.4 as a supernova remnant (SNR). This remnant is seen in projection 
over the northern border of HESS J1858+020. \citet{green09} 
estimated an age of 30000 years old for the SNR and a distance of $\sim$ 10.5 kpc  
based on the proximity of the remnant with the HII region G35.5-0.0.
As pointed out in several works (see \citealt{aha96,yamazaki06,gabici07,gabici09}), a SNR, in particular 
an old remnant (i.e age larger than a few 10000 years), interacting with a molecular cloud 
could explain the origin of the $\gamma-$ray emission via pion decay 
from proton-proton collisions. In this context, we expect a correlation 
between $\gamma-$ray emission and matter concentration. 

In this letter we study the interstellar 
medium around the sources G35.6-0.4 and HESS J1858+020 to investigate 
the possible origin for the very-high energy emission.

\section{Data}

To analyze the ISM towards HESS J1858+020 we used the \3 J=1--0 emission obtained from the 
Galactic Ring Survey. This survey maps the Galactic Ring with an angular and spectral resolution 
of 46\s~and 0.2 \k, respectively (see \citealt{jackson06}).
We also used the mosaiced images from
GLIMPSE and MIPSGAL and the GLIMPSE Point-Source Catalog (GPSC) in the {\it Spitzer}-IRAC (3.6, 4.5, 5.8 and 8 $\mu$m).
IRAC has an angular resolution between 1\farcs5 and 1\farcs9 (see \citealt{fazio04} and \citealt{werner04}).
MIPSGAL is a survey of the same region as GLIMPSE, using MIPS instrument (24 and 70 $\mu$m) on {\it Spitzer}.
The MIPSGAL resolution at 24 $\mu$m is 6\s.

\section{Results and Discussion}

\begin{figure}[h]
\centering
\includegraphics[width=9cm]{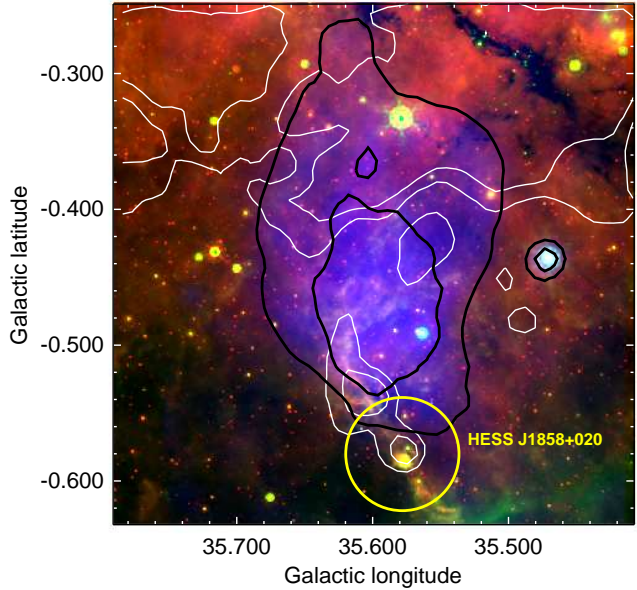}
\caption{Three color image of the analyzed region (red = 8 $\mu$m, green = 24 $\mu$m, and blue = 20 cm radio 
continuum emission). 
The black contours delineate the radio continuum emission with levels of 20 and 28 K. The white contours are 
the \3 J=1--0 emission integrated between 51 and 59 \k, its levels are 3.2 and 5 K \k. The yellow circle 
represents the source HESS J1858+020.}
\label{20cmMips8}
\end{figure}

Analyzing the whole \3 J=1--0 data cube in a region containing 
SNR G25.6-0.4 and HESS J1858+020, we find a molecular cloud 
lying in the southern border of the remnant in the velocity range between
51 and 59 \k. This velocity range coincides with the systemic velocity of G25.6-0.4 derived from
its estimated distance \citep{green09}.
Figure \ref{20cmMips8}
shows a three color image where red represents the {\it Spitzer}-IRAC 8 $\mu$m band, green is the MIPSGAL
emission at 24 $\mu$m and blue is the radio continuum emission of G35.6-0.4 at 20 cm from the VGPS. Black
contours delineate this last emission and white contours are the \3 J=1--0 emission integrated between 51 and 59 \k.
As mentioned above, the extension of the HESS source is slightly elliptical with $\sim$5\m~along 
its major axis. As the extent and angle are not well constrained, in Figure \ref{20cmMips8} we represent 
the position and the extension of HESS J1858+020 with a circle of 5\m~in diameter.
From this figure, it can be seen 
that the molecular cloud presents two well defined clumps, one, centered
at $l = 35$\fdg$60$, $b = -0$\fdg$53$, over the SNR shell, and the other, centered at 
$l = 35$\fdg$58$, $b = -0$\fdg$58$, in coincidence with the center of HESS J1858+020. 
The $\gamma-$ray emission could be related to both molecular clumps.

\begin{figure}[h]
\centering
\includegraphics[width=9cm]{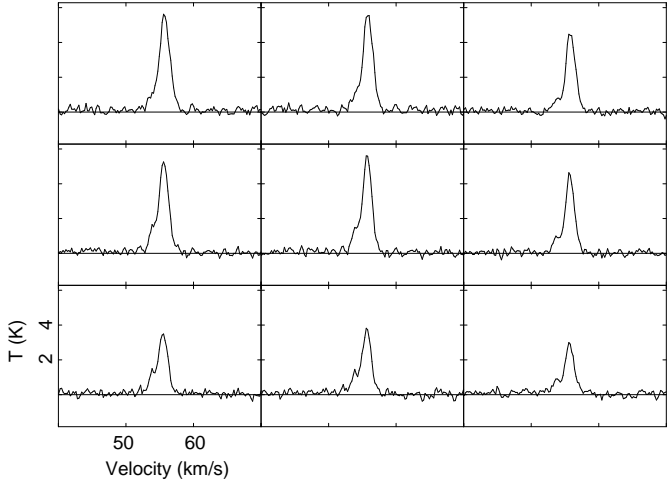}
\caption{\3 J=1--0 spectra obtained towards the maximum of the molecular clump that lies over
the SNR shell. The spectra, which show brightness temperature vs. velocity, correspond to nine positions 
observed in a region of about 50\s~$\times$ 50\s~centered at the molecular clump peak.}
\label{spectra}
\end{figure}

By inspecting the \3 spectra we found that the clump located over 
the SNR shell shows some possible kinematical evidence of shocked gas. As Fig. \ref{spectra}
displays, the spectra are not symmetric and they present a slight spectral shoulder or a less intense component at 
``blueshifted'' velocities. It could be evidence of turbulent motion in the gas, may be produced by the SNR shock
(see e.g. \citealt{falgarone94}). 
Taking into account the positional and kinematical agreement between the molecular 
gas and the SNR G35.6-0.4, we suggest that the remnant is interacting with the adjacent
molecular cloud.

\begin{figure}[h]
\centering
\includegraphics[width=9cm]{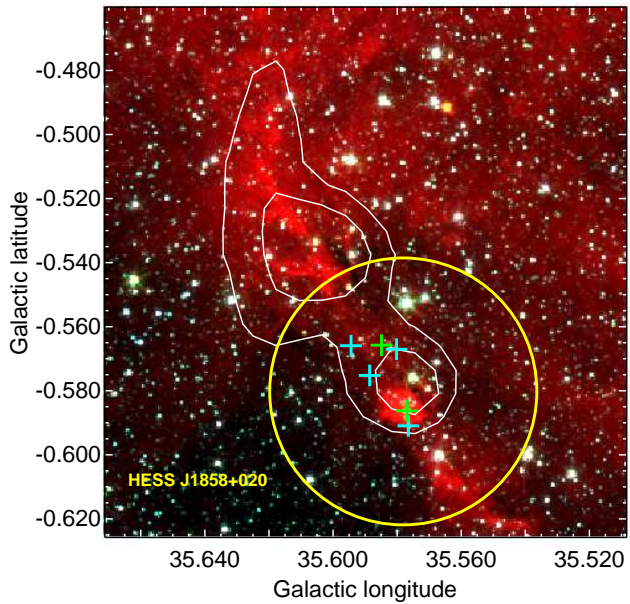}
\caption{Three color image of IRAC-{\it Spitzer} bands
(red = 8 $\mu$m, green = 4.5 $\mu$m, and blue = 3.6 $\mu$m). The white contours are
the \3 J=1--0 emission integrated between 51 and 59 \k, its levels are 3.2 and 5 K \k. The crosses are YSO
candidates according to the photometric study described in the text. Green crosses are Class I sources and cyan
crosses represents Class II sources.}
\label{3irac}
\end{figure}

Using the \3 J=1--0 line and assuming local thermodynamic equilibrium (LTE) we estimate the H$_{2}$ column density
towards the molecular clumps described above. We use:
$${\rm N(^{13}CO)} = 2.42 \times 10^{14} \frac{T_{\rm ex} \int{\tau_{13} dv}}{1 - exp(-5.29/T_{\rm ex})}$$
to obtain the \3 column density. $\tau_{13}$ is the optical depth of the line, we assume that
$T_{\rm ex}$ is 10 K and the \3 emission is optically thin. We use the relation 
N(H$_{2}$)/N(\3)$ \sim 5 \times 10^5$ (e.g. \citealt{simon01}). 
Finally, we estimate a molecular mass and a density of $\sim 5 \times 10^{3}$ \msol~and $\sim 500$ cm$^{-3}$, 
respectively, for each molecular clump. The mass value
was obtained from:
$$ {\rm M} = \mu~m_{{\rm H}} \sum{\left[ D^{2}~\Omega~{\rm N(H_{2})} \right] }, $$
where $\Omega$ is the solid angle subtended by the \3 J=1--0 beam size, $m_{\rm H}$ is the hydrogen mass,
$\mu$, the mean molecular weight, is assumed to be 2.8 by taking into account a relative helium abundance
of 25 \%, and $D$ is the distance assumed to be 10.5 kpc. Summation was performed over all the observed 
positions within the 3.5 K \ks contour level (not shown in Figures \ref{20cmMips8} and \ref{3irac}). 

We can estimate the required density matter in the $\gamma-$ray
production region for hadrons from the observed $\gamma-$ray flux of
HESS J1858+020. From $$ dF_{\rm \gamma}/dE (>E_{min}) = N_{0} (E/1 {\rm Tev})^{-\Gamma},$$ we obtain 
$F_{\rm \gamma} \sim 1.2 \times 10^{-12}$ cm$^{-2}$s$^{-1}$.
Here $E_{min} = 0.5$ TeV, $\Gamma = 2.17$, $N_{0} = 0.6 \times 10^{-12}$ cm$^{-2}$s$^{-1}$ \citep{aha08}. 
Using the equation 16 in \citet{torres03}, we obtain a density of about 150 cm$^{-3}$ assuming
an acceleration efficiency of hadrons of the order of 3$\%$ and a supernova power
of 10$^{51}$ ergs.
Thus the molecular gas in positionally coincidence with HESS J1858+020 is
densely enough to generate the very-high energy emission hadronically.

Regarding the IR emission in the studied region, from Fig. \ref{20cmMips8}
it can be seen the presence of at least two bright sources  embedded
in the southern molecular clump. Also the mid-IR emission shows the presence of hot dust and 
policyclic aromatic hydrocarbons (PAHs), green and red in Fig. \ref{20cmMips8}, respectively, 
suggesting an active star forming region.
In order to confirm that, we carried out an IR photometric study of 
the sources that lie over the molecular clump that is centered at the HESS source position.

In Figure \ref{3irac} we present a three color image with three IRAC-{\it Spitzer} bands: red is 8 $\mu$m, 
green is 4.5 $\mu$m and blue is 3.6 $\mu$m. As in Figure \ref{20cmMips8} the white contours are the \3 J=1--0 
emission integrated between 51 and 59 \k.
To look for tracers of star formation activity,
we use the GLIMPSE Point-Source Catalog to perform photometry.
Considering only sources that have been detected in the four {\it Spitzer}-IRAC bands, we found 32 sources 
within a circular area of 80\s~in radius centered at the molecular clump. According to the criteria 
presented by \citet{all04}, we found 6 sources as young stellar objects (YSO)
 candidates. Two of them are Class I objects 
(protostar with circumstellar envelope; green crosses in Fig. \ref{3irac}) and the others are 
Class II objects (young stars with only disk emission; cyan crosses in Fig. \ref{3irac}).  From Fig. 3 it can be noticed a bright source  
that lies almost at the center of the molecular clump, but in this case since 
it has some bands with null fluxes in the GLIMPSE 
catalog, we can not perform photometry. However, from Fig. \ref{3irac} can be appreciated 
that this source appears slightly extended in the 4.5 $\mu$m emission (green), which according to 
\citet{cyga08}, may suggest that it could be a MYSO driving outflows.
For the sources that have also 
detection in the 2MASS {\it JHK}-bands, we derive their spectral energy distribution (SED)
by fitting the fluxes using the tool developed by
\citet{robit07} and available online\footnote{http://caravan.astro.wisc.edu/protostars/}. This was 
possible only for three Class II sources, resulting that they are indeed 
young objects ($\sim 10^{5}$ years).
We conclude that this region is active in star formation, which can be considered as an alternative or 
complementary possibility in order to explain the very-high energy emission. 
HESS J1858+020 could be a similar case as SNR W28. Several molecular clumps and star-forming 
regions were found towards W28 and proposed to be candidates to explain the very-high energy emission detected in 
the region \citep{aha08b}. Further sub-mm observations, high CO transitions and  
tracers of star formation, will provide more accurate mass and density estimates and allow to search for 
perturbed gas by both SNR shock and star formation processes.

\section{Summary }

In this work we investigate the ISM towards the southern border of the recently re-identified source 
G35.6-0.4 as a SNR, which coincides with the northern portion of the very-high energy source HESS J1858+020.
Analyzing the \3 J=1--0 emission we find a molecular cloud in positional and kinematical agreement 
with the SNR, suggesting that the molecular gas is being affected by the remnant shock.
The discovered molecular cloud is composed by two clumps, each with mass and density of 
$\sim 5 \times 10^{3}$ \msol~and $\sim 500$ cm$^{-3}$, respectively. 
Considering the gamma-ray flux observed towards HESS J1858+020, 
we estimate that a molecular cloud with a density of at least 150 cm$^{-3}$ could explain the 
very-high energy emission hadronically. Thus, we suggest that the interaction between
the SNR G35.6-0.4 and the molecular cloud is responsible for the $\gamma-$ray emission.
Additionally, analyzing mid-IR emission, we find that the region is active in star formation, which
could be considered as an alternative or
complementary possibility to explain the very-high energy emission. This SNR -- molecular cloud -- HESS source
complex could be a similar case as the high-energy sources detected towards SNR W28.
Further sub-mm observations will provide more accurate mass and density estimates and allow to search for
perturbed gas by both SNR shock and star formation processes.

\begin{acknowledgements}

S.P. and E.G. are members of the {\sl Carrera del
investigador cient\'\i fico} of CONICET, Argentina. 
This work was partially supported by the CONICET grant PIP 112-200801-02166, UBACYT A023 and ANPCYT
PICT-2007-00902.

\end{acknowledgements}

\bibliographystyle{aa}  
\bibliography{bibHess}
\IfFileExists{\jobname.bbl}{}
{\typeout{}
\typeout{****************************************************}
\typeout{****************************************************}
\typeout{** Please run "bibtex \jobname" to optain}
\typeout{** the bibliography and then re-run LaTeX}
\typeout{** twice to fix the references!}
\typeout{****************************************************}
\typeout{****************************************************}
\typeout{}
}

\end{document}